\documentclass[12pt,final]{amsart}
\newcommand{\normalspacing}{\renewcommand{\baselinestretch}{1.1}\tiny\normalsize}

\normalspacing

\usepackage{amssymb,alltt,verbatim,xspace}
\usepackage{hyperref}
  \usepackage[final,dvips]{graphicx}

\theoremstyle{plain}
\newtheorem*{thm*}{Theorem}

\theoremstyle{definition}

\theoremstyle{remark}

\newcommand{\ddt}[1]{\ensuremath{\frac{\partial #1}{\partial t}}}

\newcommand{\ddz}[1]{\ensuremath{\frac{\partial #1}{\partial z}}}
\newcommand{\ddzz}[1]{\ensuremath{\frac{\partial^2 #1}{\partial z^2}}}

\newcommand{\ip}[2]{\ensuremath{\left<#1,#2\right>}}

\newcommand\lam{\lambda}

\newcommand{\Tpmp}{T_{\text{pmp}}}

\begin{document}
\title[Exact temperatures in ice and bedrock]{An exact solution to the temperature equation \\ in a column of ice and bedrock}

\author{Ed Bueler}

\date{\today.   Dept.~of Mathematics and Statistics, Univ.~of Alaska, Fairbanks.}

\maketitle
\thispagestyle{empty}

\section{The problem} The goal here is fairly straightforward.  We want a solution of a pure conduction problem in ice \emph{and} bedrock.  This solution needs to be suitable for verifying a numerical scheme for conservation of energy.  This solution will also help with the construction of an approximate polythermal scheme.  We will use this exact solution in the context of a coupled ice flow and conservation of energy model, namely PISM \cite{pism-web-page}.  This exact solution will form one of a suite of verification tests for PISM \cite{BB,BBL,BLKfastearth,BLKCB}.

In particular, we will find a function $T(z,t)$ with the following properties
\begin{gather*}
T(H,t) = T_s, \\
\rho_I c_I \ddt{T} = k_I \ddzz{T} \qquad (0 < z < H), \\
T(0^+,t) = T(0^-,t), \\
k_I \ddz{T}(0^+,t) = k_R \ddz{T}(0^-,t), \\
\rho_R c_R \ddt{T} = k_R \ddzz{T} \qquad (-B < z < 0), \\
-k_R \ddz{T}(-B,t) = G.
\end{gather*}
The two conditions at the ice/rock interface $z=0$ are continuity of temperature and of heat flux, respectively.

The ice thickness is $H>0$ and the bed thickness is $B>0$; representative values used here are
	$$B=1000\, \text{m} \quad \text{and} \quad H = 3000\, \text{m}.$$
The ice occupies $0<z<H$ and has density $\rho_I$, specific heat capacity $c_I$, and conductivity $k_I$.  The bedrock occupies $-B<z<0$ and has density $\rho_R$, specific heat capacity $c_R$, and conductivity $k_R$.  Reasonable values of these constants are given in the C implementation at the end.  The constant value $T_s$ of the surface temperature will be $223.15$ K or $-50\,\phantom{|}^\circ\text{C}$.  The value of the geothermal flux used here is
	$$G = 42\,\text{mW}/\text{m}^2.$$

\begin{figure}[ht]
\includegraphics[height=3.2in,keepaspectratio=true]{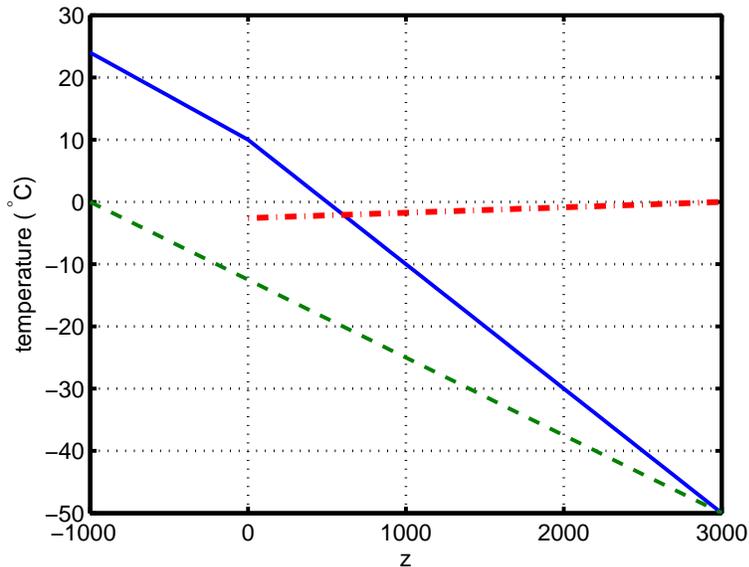}
\caption{Final temperature $T(z,+\infty)$ (solid) and initial temperature $T(z,0)$ (dashed).  Within the ice the temperature must actually remain below the pressure-melting temperature $\Tpmp(z)$ (dash-dotted).}
\label{fig:initialfinal}
\end{figure}

Let us take as our initial condition an (absolute) temperature which is a linearly-increasing function of the depth below the surface of the ice:
\begin{equation}\label{initialcond}
T(z,0) = T_s + \phi (H - z), \qquad T_s = 223.15\, \text{K}, \qquad \phi 
       = 0.0125\, \text{K}\,\text{m}^{-1}
\end{equation}
Figure \ref{fig:initialfinal} includes a graph of this linear initial condition, which warms from $-50\phantom{|}^\circ\text{C}$ at the surface to $0\phantom{|}^\circ\text{C}$ at the base of the bedrock layer (i.e.~at depth $1000$ m into the bedrock).

In fact we will slightly revise this initial condition in Section \ref{sect:testKsoln}.  In particular, for numerical accuracy reasons, it will be desirable to use an initial condition with a finite eigenfunction expansion.  The graph of the initial condition in Figure \ref{fig:initialfinal} is accurate at printer resolution, however.

As noted, a goal is to verify parts of the thermomechanical model in PISM.  On the other hand, PISM is primarily a three (spatial) dimensional model for the flow of ice, coupled with the thermodynamics of the ice and the bedrock.  Therefore, in using the exact temperature solution here for verification, we will suppose that the conditions for the full, coupled model are ice of constant thickness $H$ everywhere, accumulation which is identically zero, and a flat bed.  Then PISM will predict no flow.  In particular, the advection, strain-heating, and basal frictional heating parts of the general conservation of energy equations are each identically zero.  So in this case we see that the temperature problem above is all that remains to solve in the full, coupled model.  Note that other tests fully verify the conservation of energy numerical scheme in flowing ice \cite{BB,BBL}, but they do not include heat storage in bedrock.

We have not included melting in the above.  Recall that the pressure-melting temperature in the ice is $\Tpmp(z) = T_0 - \beta (H-z)$.  (We suppose $T_0 = 273.15$ K and $\beta = 8.66\times 10^{-4}\,\text{K}/\text{m}$ for concreteness.  With these constants $\Tpmp(0) = -2.598\,\phantom{l}^\circ\text{C}$.)  For sufficiently large $t>0$, the solution to the above problem has $T(z,t) > \Tpmp(z)$ for some locations $z\ge 0$.  At such locations the above model no longer applies because there will be partial melting in the ice.  For verification purposes we are interested in the first time at which melting occurs.

\section{Finding an eigenfunction expansion}

We find a classical kind of solution to this classical kind of problem.  First we transform our inhomogeneous problem to a homogeneous one.  Let
\begin{equation}\label{Pdefn}
  P(z) = \begin{cases} z/k_I - H/k_I, &0 \le z \le H \\ z/k_R - H/k_I, &-B \le z \le 0. \end{cases}
\end{equation}
Define the rescaled temperature
    $$\theta(z,t) = T(z,t) - T_s + G P(z).$$
It is straightforward to check that $T(z,t)$ solves the original problem if and only if $\theta(z,t)$ solves
\begin{gather*}
\theta(H,t) = 0, \\
\rho_I c_I \theta_t = k_I \theta_{zz} \qquad (0 < z < H), \\
\theta(0^+,t) = \theta(0^-,t), \\
k_I \theta_z(0^+,t) = k_R \theta_z(0^-,t), \\
\rho_R c_R \theta_t = k_R \theta_{zz} \qquad (-B < z < 0), \\
\theta_z(-B,t) = 0.
\end{gather*}
This boundary value problem is linear and homogeneous.  (Note we have also switched to subscript notation for derivatives.)

The rescaled temperature $\theta$ has initial condition 
\begin{equation}\label{thetainitcond}
\theta(z,0) = G P(z) + \phi (H-z).
\end{equation}

We expect that the above problem for $\theta(z,t)$ is well-posed \cite{Evans}.  Furthermore we expect that $\lim_{t\to+\infty} \theta(z,t)=0$, that is, we expect that the problem is asymptotically stable.  Thus we expect
    $$T(z,+\infty) = T_s - G P(z).$$
Of course this would violate the requirement that $T \le \Tpmp$ within the ice.  Nonetheless this final state is worth graphing along with the initial state and the pressure-melting temperature, as in Figure \ref{fig:initialfinal}.

Next we separate variables and seek eigenfunctions.  Preliminary thoughts might go like this: If $\theta(z,t)=f_I(z) g(t)$ on the interval $0<z<H$ and $\theta(z,t)=f_R(z) g(t)$ on the interval $-B<z<0$ then we have
    $$\rho_I c_I \frac{\dot g}{g} = k_I \frac{f_I''}{f_I} \quad \text{ and } \quad \rho_R c_R \frac{\dot g}{g} = k_R \frac{f_R''}{f_R}.$$
This separated form for $\theta(z,t)$ must have, and does have, the same dependence on $t$ in both the ice and the bedrock.  The solution must satisfy boundary conditions at $z=H$ and $z=-B$.  As usual for the heat equation, the solution decays exponentially in time and is (roughly) sinusoidal in space.  The conditions at $z=0$ correspond to continuity of the solution and of the heat flux, and this means a change in amplitude for the sinusoid because the conductivity changes.

A conclusion to the above thoughts is an \emph{ansatz} for separated solutions:\footnote{The reader who does not like this language may confirm that, at the end, we have a full spectral resolution of our discrete spectrum, self-adjoint operator.}
\begin{equation}\label{ansatz}
\theta(z,t) = e^{-\lam t}\,\begin{cases} \sin(\alpha(H-z)), &0< z < H, \\ \gamma \cos(\beta(B+z)), &-B < z < 0.\end{cases}
\end{equation}
The eigenvalues are denoted $\lam$.  We will see that they form a countable sequence $0 < \lam_0 < \lam_1 < \lam_2 < \dots$ which tends to positive infinity.  The eigenfunctions are the spatial parts of the corresponding \emph{ansatz} solutions.

The constants $\lam,\alpha,\beta,\gamma$ are determined by the connection conditions and the PDEs themselves:
\begin{align}
\rho_I c_I \lam &= k_I \alpha^2, \label{cond1}\\
\sin(\alpha H) &= \gamma \cos(\beta B), \label{cond2}\\
\alpha k_I \cos(\alpha H) &= \beta \gamma k_R \sin(\beta B), \label{cond3} \\
\rho_R c_R \lam &= k_R \beta^2. \label{cond4}
\end{align}
Conditions \eqref{cond1} and \eqref{cond4} combine to eliminate $\lam$ and give
\begin{equation}\label{betafromalpha}
\beta = Z\,\alpha
\end{equation}
where
    $$Z=\sqrt{\frac{\rho_R c_R\, k_I}{k_R\, \rho_I c_I}}.$$
On the other hand, conditions \eqref{cond2} and \eqref{cond3} combine to eliminate $\gamma$.  Indeed, using \eqref{betafromalpha} as well, after clearing fractions one gets
\begin{equation} \label{compatible}
A \sin(H\alpha) \sin(Z B \alpha) = \cos(H\alpha) \cos(Z B \alpha)
\end{equation}
where
    $$A = \frac{k_R}{k_I} Z.$$
Using trigonometric identities one can rewrite \eqref{compatible} as
\begin{equation} \label{coscross}
\left(\frac{A-1}{A+1}\right)\, \cos((H - Z B) \alpha) = \cos((H + Z B) \alpha).
\end{equation}

Note that $0<A<1$ for reasonable values of density, specific heat capacity, and conductivity for ice and bedrock.  Thus,
    $$\left|\frac{A-1}{A+1}\right| < 1,$$
so \eqref{coscross} equates two sinusoidal functions, with the left-hand function of smaller magnitude and lower frequency.

\begin{figure}[ht]
\includegraphics[height=3.0in,keepaspectratio=true]{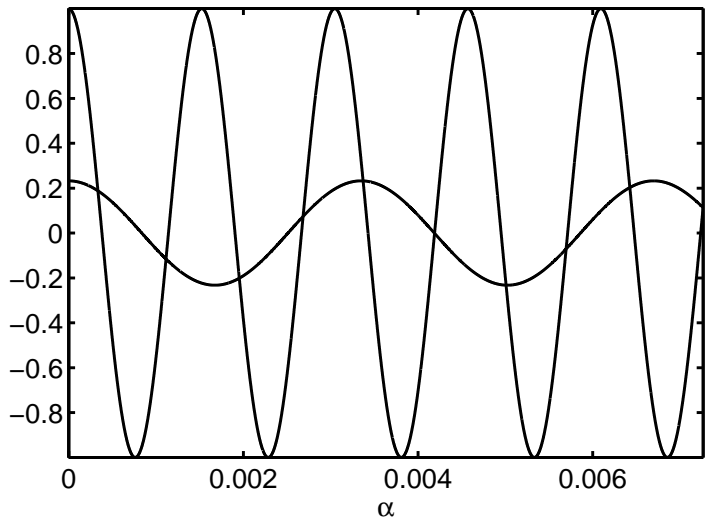}
\caption{A picture of equation \eqref{coscross}.  There is exactly one solution $\alpha_k > 0$ per half-cycle of the higher amplitude cosine.}
\label{fig:coscross}
\end{figure}

We have arrived at a visualizable stage.  Equation \eqref{coscross} determines countably many discrete values $\alpha=\alpha_k>0$, $k=0,1,2,\dots$, as shown in Figure \ref{fig:coscross}.  Solutions of \eqref{coscross} must occur between each consecutive extrema of the higher amplitude and higher frequency cosine on the right side of the equation.  Though \eqref{coscross} is transcendental, accurate solutions are easily found by good numerical methods like Brent's method \cite{BurdenFaires}.  In particular, one can bracket each solution, and Brent's method maintains such a bracket as it converges to a root.

We need only find positive solutions $\alpha$ of equation \eqref{coscross}.  They will form a positive increasing sequence $0 < \alpha_0 < \alpha_1 < \alpha_2 < \dots$.  Also, as a special case which may be used to check formulas, note that if the material constants $\rho$, $c$, and $k$ are non-physically assumed to be the \emph{same} for ice and for bedrock then $Z=1$ and $A=1$ so the equation we solve is just $\cos((H+B) \alpha) = 0$.  In this case $\alpha_k = ((2k+1) \pi)/(2(H+B))$.

\begin{figure}[ht]
\includegraphics[height=3.2in,keepaspectratio=true]{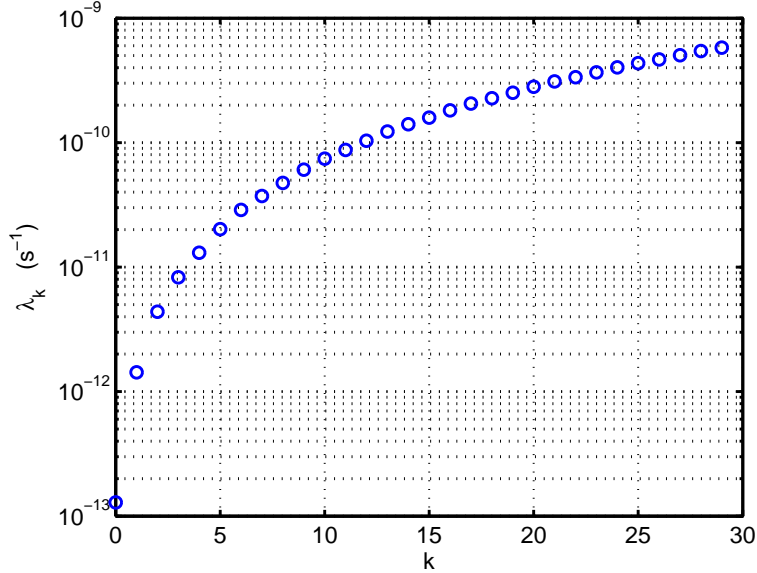}
\caption{The first 30 eigenvalues $\lambda_k$, $k=0,1,\dots,29$.}
\label{fig:spectrum}
\end{figure}

Once we $\alpha_k$ then from \eqref{betafromalpha} we get a corresponding sequence $\beta_k$.  From \eqref{cond2} or \eqref{cond3} we find $\gamma_k$.  From \eqref{cond1} or \eqref{cond4} we get the (positive) eigenvalues $\lam_k$ themselves.  In fact, using the constants specified below in the C implementation, we get the spectrum $\{\lambda_k\}$ shown in Figure \ref{fig:spectrum}.

\begin{figure}[ht]
\includegraphics[height=3.2in,keepaspectratio=true]{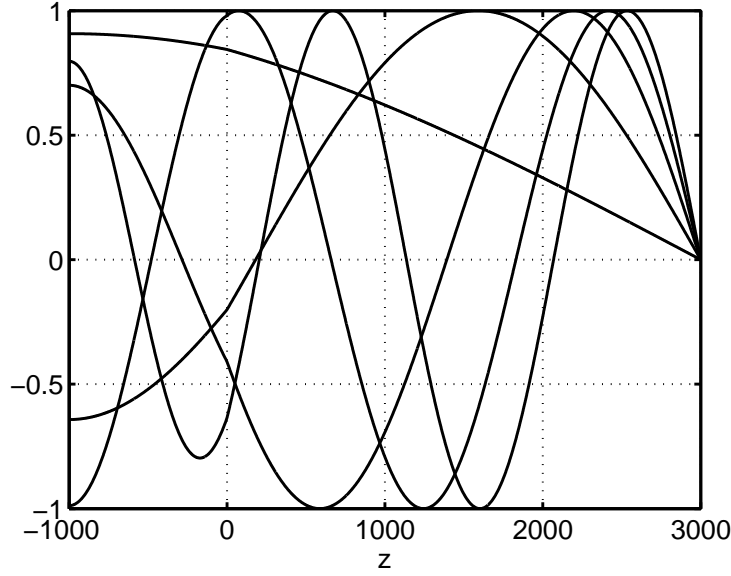}
\caption{Unnormalized eigenfunctions $\tilde \theta_0(z),\dots,\tilde \theta_4(z)$.  Note a change in amplitude at $z=0$.}
\label{fig:firstfive}
\end{figure}

The unnormalized eigenfunctions are
    $$\tilde \theta_k(z) = \begin{cases} \sin(\alpha_k(H-z)), &0< z < H, \\ \gamma_k \cos(\beta_k (B+z)), &-B < z < 0.\end{cases}$$ 
Those $\tilde\theta_k$ corresponding to the five smallest (most important) eigenvalues $\lam_k$ are shown in Figure \ref{fig:firstfive}.

The eigenfunctions $\tilde\theta_k(z)$ are solutions of a Sturm-Liouville problem  \cite{BrownChurchill}.  Thus they are an orthogonal set with respect to an appropriate inner product.  This inner product includes the coefficients used in computing the thermal energy.  In fact, recall that $\iiint_V \rho\, c\, T\,\, dx\,dy\,dz$ is the internal (specific) heat energy stored in a material with temperature $T$ occupying a volume $V$.  So, if $f(z),g(z)$ are integrable functions on $-B < z < H$, we define the inner product:
\begin{equation}\label{ipdefn}
\ip{f}{g} := \rho_R c_R \int_{-B}^0 f(z) g(z)\,dz + \rho_I c_I \int_0^H f(z) g(z)\,dz.
\end{equation}

An easy calculation computes inner products of the (as yet) unnormalized eigenfunctions, as follows.  First we transform to doable integrals,
\begin{align*}
\ip{\tilde\theta_k}{\tilde\theta_l} &= \rho_R c_R \gamma_k \gamma_l \int_{-B}^0 \cos(\beta_k (B+z)) \cos(\beta_l (B+z))\,dz \\
    &\qquad\qquad+ \rho_I c_I \int_0^H \sin(\alpha_k (H-z)) \sin(\alpha_l (H-z))\,dz \\
 &= \rho_R c_R \gamma_k \gamma_l \int_0^B \cos(\beta_k x) \cos(\beta_l x)\,dx + \rho_I c_I \int_0^H \sin(\alpha_k y) \sin(\alpha_l y)\,dy.
\end{align*}
Now there are two cases.  If $k=l$ then we have a formula for normalization constants:
\begin{align*}
X_k^2 &:= \ip{\tilde\theta_k}{\tilde\theta_k} = \frac{1}{2} \rho_R c_R \gamma_k^2 \int_0^B 1 + \cos(2 \beta_k x)\,dx + \frac{1}{2} \rho_I c_I \int_0^H 1 - \cos(2 \alpha_k y)\,dy \\
    &= \frac{1}{2} \rho_R c_R \gamma_k^2 \left(B + \frac{\sin(2 \beta_k B)}{2\beta_k}\right) + \frac{1}{2} \rho_I c_I \left(H - \frac{\sin(2 \alpha_k H)}{2\alpha_k}\right) \\ 
    &= \frac{1}{2} \left(\rho_R c_R \gamma_k^2 B + \rho_I c_I H\right) + \frac{1}{2\beta_k}\, \rho_R c_R \gamma_k^2 \sin(\beta_k B) \cos(\beta_k B) \\
    &\qquad\qquad - \frac{1}{2\alpha_k}\, \rho_I c_I \sin(\alpha_k H) \cos(\alpha_k H).
\end{align*}
This expression simplifies further using the properties of the eigenfunctions:
\begin{align*}
X_k^2 &\stackrel{\ast}{=} \frac{1}{2} \left(\rho_R c_R \gamma_k^2 B + \rho_I c_I H\right) + \frac{1}{2\beta_k}\, \rho_R c_R \gamma_k^2 \sin(\beta_k B) \cos(\beta_k B) \\
    &\qquad\qquad - \frac{1}{2\alpha_k}\, \rho_I c_I \gamma_k \cos(\beta_k B) \frac{\beta_k\gamma_k k_R}{\alpha_k k_I} \sin(\beta_k B) \\
    &= \frac{1}{2} \left(\rho_R c_R \gamma_k^2 B + \rho_I c_I H\right) + \frac{\gamma_k^2}{2\beta_k\alpha_k^2 k_I}\,\sin(\beta_k B) \cos(\beta_k B) \left(\rho_R c_R \alpha_k^2 k_I - \rho_I c_I \beta_k^2 k_R\right) \\
    &\stackrel{\ast\ast}{=} \frac{1}{2} \left(\rho_R c_R \gamma_k^2 B + \rho_I c_I H\right).
\end{align*}
The starred equality follows from equations \eqref{cond2} and \eqref{cond3}.  The double-starred equality follows from equations \eqref{cond1} and \eqref{cond4}.

If $k\ne l$ we get $\ip{\tilde\theta_k}{\tilde\theta_l}=0$, but we omit the details.

Thus the \emph{normalized} eigenfunctions are 
	$$\theta_k(z) = \frac{\tilde\theta_k(z)}{X_k} = \frac{1}{X_k} \begin{cases} \sin(\alpha_k(H-z)), &0< z < H, \\ \gamma_k \cos(\beta_k (B+z)), &-B < z < 0.\end{cases}$$

\section{The solution to the time-dependent problem}\label{sect:testKsoln}

The solution to the time-dependent problem for $\theta(z,t)$ is the infinite series
\begin{equation}\label{thetasum}
\theta(z,t) = \sum_{k=0}^\infty C_k e^{-\lambda_k t} \theta_k(z)
\end{equation}
where $\theta_k(z)$ are the normalized eigenfunctions computed above, and $C_k = \ip{\theta_k}{\theta(t\!=\!0)}$.  In fact,
	$$C_k = \rho_I c_I \int_0^H \theta_k(z) \left(G P(z) + \phi (H-z)\right)\,dz + \rho_R c_R \int_{-B}^0 \theta_k(z) \left(G P(z) + \phi (H-z)\right)\,dz$$
from equation \eqref{thetainitcond}.  Also note $P(z)$ is given in equation \eqref{Pdefn}.  We can naturally describe $C_k$ as a linear combination of definite integrals:
\begin{equation}\label{Cksplit}
C_k = X_k^{-1}\, \left(\rho_I c_I\, I_k^1 + \rho_R c_R \gamma_k\, I_k^2\right),
\end{equation}
where
\begin{align*}
I_k^1 &= \int_0^H \sin(\alpha_k(H-z))\,\left(G\left(\frac{z}{k_I} - \frac{H}{k_I}\right) + \phi(H-z)\right)\,dz
\end{align*}
and
\begin{align*}
I_k^2 &= \int_{-B}^0 \cos(\beta_k(B+z))\,\left(G\left(\frac{z}{k_R} - \frac{H}{k_I}\right) + \phi(H-z)\right)\,dz.
\end{align*}
These are elementary integrals, though it is easy to get things wrong anyway.   They simplify to
\begin{equation}
I_k^1 = -\left(\frac{G}{k_I} - \phi\right)\,\alpha_k^{-2}\left[\sin(\alpha_k H) - (\alpha_k H)\cos(\alpha_k H)\right],\label{I1final}
\end{equation}
\begin{align}
I_k^2 &= \left(\frac{G}{k_R} - \phi\right)\,\beta_k^{-2}\left[\cos(\beta_k B) - 1 + (\beta_k B)\sin(\beta_k B)\right]\label{I2final} \\
&\quad - \left(B\left(\frac{G}{k_R} - \phi\right) + H \left(\frac{G}{k_I} - \phi\right)\right)\,\beta_k^{-1}\sin(\beta_k B).\notag
\end{align}
The temperature itself (not rescaled) is given by
\begin{equation}\label{Tfromtheta}
T(z,t) = \theta(z,t) + T_s - G P(z).
\end{equation}
Formulas \eqref{thetasum}, \eqref{Cksplit}, \eqref{I1final}, \eqref{I2final}, and \eqref{Tfromtheta} together form the time-dependent solution to the initial value problem specified so far.

Now, the infinite sum converges quickly for large times but it converges rather slowly for $t=0$.  This fact relates to the poor differentiability of the initial state (times the diffusivity, that is), and it is a common situation for conduction problems \cite{BrownChurchill}.  To avoid any concern with convergence at $t=0$, we \emph{redefine the initial state to have a finite eigenfunction expansion}.  That is, we replace equation \eqref{thetainitcond} with the revised condition
\begin{equation}\label{revised_thetainitcond}
\theta(z,0) = \sum_{k=0}^{29} C_k \theta_k(z)
\end{equation}
where the coefficients $C_k$ are given exactly as before by equations \eqref{Cksplit}, \eqref{I1final}, and \eqref{I2final}.  This represents a change of the initial condition by a maximum of only about $0.001$ K, so to printer or screen accuracy this is not important, and indeed the upper limit of the sum $N=29$ was chosen for such reasons.  But that detail is not important.  Rather, the point is that by making this change any concerns about evaluating the exact solution to high accuracy are immediately resolved, and this is our goal.  Note that this change also means that the time-dependent solution has a finite expansion:
\begin{equation}\label{Tfiniteexpansion}
T(z,t) = T_s - G P(z) + \sum_{k=0}^{29} C_k e^{-\lambda_k t} \theta_k(z).
\end{equation}

\section{Verification of PISM using this exact solution}

The exact solution given by equation \eqref{Tfiniteexpansion} is verification Test \textbf{K} in PISM \cite{pism-web-page}.  As previously noted, PISM is a three-dimensional ice flow simulation program which includes many coupled physical models.  Here we use Test \textbf{K} to verify the part of PISM which relates to the simulation of heat conduction.  That is, PISM contains a semi-implicit finite difference approximation of a shallow (continuum) approximation of the conservation of energy equation.  Our use of Test \textbf{K} for verification concerns only the pure conduction aspect of that scheme.

We note that bugs can and have appeared in the part of PISM which numerically approximates the point in the bedrock where the geothermal flux is applied and at the switch of material properties from ice to bedrock.  Of course in a many-physical-models code like PISM there are many contributions to the approximation of conservation of energy at the ice-bedrock interface, including basal melting and frictional heating, and thus the numerical scheme for grid points at the base of the ice is complicated.  An exact solution is helpful for debugging such details even if it only verifies a sub-model of the full ``multi-physics''.

The vertical grid in PISM has, for now, constant spacing $\Delta z$, and indeed this spacing is equal in both the ice and the bedrock.\footnote{This statement applies to PISM in October 2007, but future versions may be change.  Such changes to the grid are exactly the kind of numerical issue which motivates building and documenting a suite of exact solutions for verification.}  

The numerical scheme in PISM for the energy equation is documented in the Appendices of \cite{BBL}.  The scheme is semi-implicit generally, but when restricted to pure conduction in a column of ice, as here, it is fully-implicit.  That is, it corresponds to centered-spatial-differencing and backward Euler method in time and thus it has local truncation error $O(\Delta t, \Delta z^2)$.  It is unconditionally stable (for pure conduction).  Indeed, the numerical issues associated to advection and to strain heating, as discussed in \cite{BBL}, are not important here.

To verify using Test \textbf{K} we choose a refinement path \cite{MortonMayers} with $\Delta z = 100, 50, 25, 12.5$, and $6.25$ meters.  As long as $\Delta t$ is reduced appropriately, which means $\Delta t = C \Delta z^2$ for some appropriate $C$, this gives a refinement path along which the error should decay by a factor of four at each refinement.   In fact we use $\Delta t = 400, 100, 25, 6.25$, and $1.5625$ years, so in fact $C=0.04$.  Because the exact and numerical solutions have constant dependence on $x$ and $y$, the horizontal grid is fixed as at a convenient (very coarse) level.

\begin{figure}[ht]
\includegraphics[height=3.0in,keepaspectratio=true]{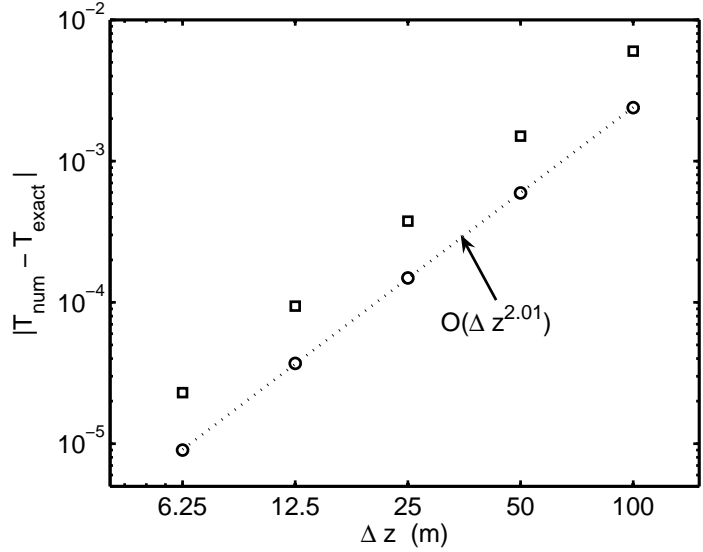}
\caption{Maximum (squares) and average (circles) errors made by PISM in approximating the temperature \emph{within the ice} in Test \textbf{K}.}
\label{fig:iceTconv}
\end{figure}

\begin{figure}[ht]
\includegraphics[height=3.0in,keepaspectratio=true]{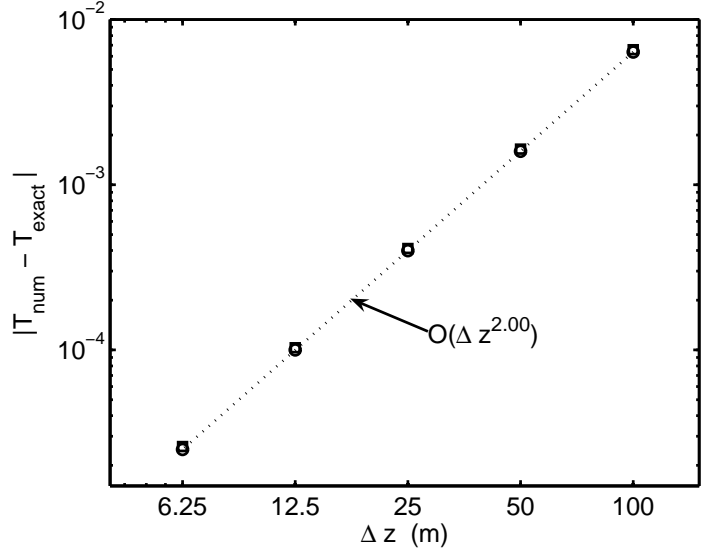}
\caption{Maximum (squares) and average (circles) errors made by PISM in approximating the temperature \emph{within the bedrock} in Test \textbf{K}.}
\label{fig:rockTconv}
\end{figure}

As shown in Figures \ref{fig:iceTconv} and \ref{fig:rockTconv}, which are admittedly boring figures, the maximum and average numerical errors at all points within the ice and within the bedrock do decay to zero.\footnote{Mathematical readers should note that we are reporting both $L^\infty$ and $L^1$ error.}  As shown in the figures, fitting the average error versus $\Delta z$ to a curve of the form $(\text{err}) = A (\Delta z)^r$ gives $r=2.01$ for the approximation within the ice and $r=2.00$ within the bedrock.

This suggests that the numerical scheme is achieving the optimal rate, that is, the local truncation error is reflected in the global approximation error.

\begin{figure}[ht]
\includegraphics[height=3.0in,keepaspectratio=true]{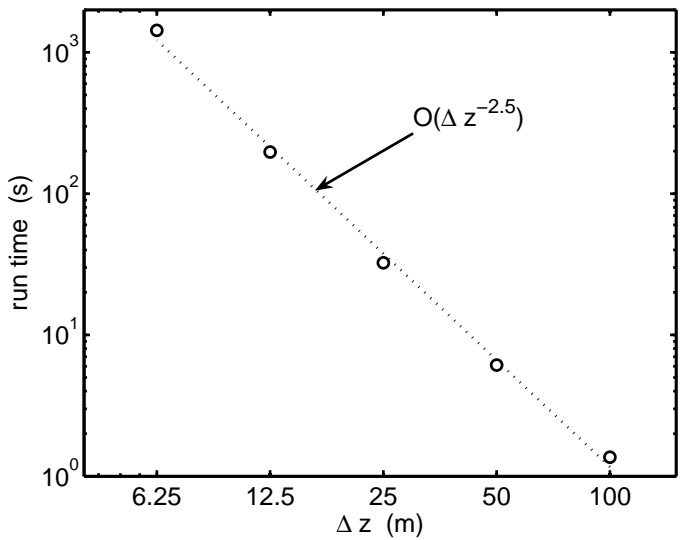}
\caption{Run time for PISM to complete Test \textbf{K} using two processors.}
\label{fig:timing}
\end{figure}

Note that along this refinement path, as $\Delta z$ is reduced by a factor of two we must reduce $\Delta t$ by a factor of four if we want the time part of the local truncation error to contribute a comparable fraction of the error.  Along this refinement path the amount of computational work \emph{per step} therefore goes up by a factor of two but the amount of computational work \emph{per model year} goes up by a factor of eight.  This statement turns out to be slightly pessimistic, because Figure \ref{fig:timing} suggests that, running in parallel with two processors, the run time for PISM is related to the $-2.5$ power of $\Delta z$.  That is, instead of a halving of $\Delta z$ generating a slowdown of a factor of $8=2^3$, there seems to be a slowdown by a factor of only $2^{2.5}$.  This is probably related to the increasing efficiency of the code as more points are computed in each column.

Finally, the critical time $t$ when $T(0,t)$ first exceeds pressure-melting is between $133,000$ years and $134,000$ years.  Indeed, by bisection on the exact solution, it must be within a year of $133,465$ years.  With a modestly refined grid with $\Delta z = 25$ m we see the numerical approximation first has $T(0,1)$ reach pressure melting between $133,470$ and $133,480$ model years.  This seems close enough, and no further verification has been pursued.


\appendix
\section{Reference implementation of Test \textbf{K}}

This Appendix contains a C code which accepts $t$ and $z$ and computes the (absolute) temperature $T$ given by equation \eqref{Tfiniteexpansion}.  That is, this code evaluates Test \textbf{K} in PISM.  It has only been compiled with the GNU \verb|gcc| compiler, and the reader may note that it is not particularly written for efficiency or speed.

The file which contains the code is called \verb|exactTestK.c|, and it is listed verbatim.  A header file \verb|exactTestK.h| exists in the PISM source tree, but listing it here would add no information so it is omitted.  Likewise there is also a simple example program \verb|simpleK.c| for evaluating the exact solution, but we do not list it.

The procedure \verb|exactK()| in \verb|exactTestK.c| is devoted to evaluating the exact solution using saved values of $\alpha_k$.  These values may be recomputed using the part of the code which is delimited by ``\verb|#if COMPUTE_ALPHA|'' and ``\verb|#endif|''.  This latter part uses Brent's method, as implemented in the \href{http://www.gnu.org/software/gsl/}{GNU Scientific Library}, to solve equation \eqref{coscross} numerically to about 14 digits of accuracy (in double precision).

The numerical approximation of conservation of energy within PISM is, of course, not listed here.  The latest revision can be found at the PISM download site \cite{pism-web-page}.

\tiny
\bigskip
\centerline{\rule{4.6in}{0.5mm}}
\smallskip
\begin{verbatim}
/*
   Copyright (C) 2007 Ed Bueler
  
   This file is part of PISM.
  
   PISM is free software; you can redistribute it and/or modify it under the
   terms of the GNU General Public License as published by the Free Software
   Foundation; either version 2 of the License, or (at your option) any later
   version.
  
   PISM is distributed in the hope that it will be useful, but WITHOUT ANY
   WARRANTY; without even the implied warranty of MERCHANTABILITY or FITNESS
   FOR A PARTICULAR PURPOSE.  See the GNU General Public License for more
   details.
  
   You should have received a copy of the GNU General Public License
   along with PISM; if not, write to the Free Software
   Foundation, Inc., 51 Franklin St, Fifth Floor, Boston, MA  02110-1301  USA
*/

#include <stdio.h>
#include <math.h>
#include <gsl/gsl_errno.h>
#include <gsl/gsl_math.h>
#include <gsl/gsl_roots.h>
#include "exactTestK.h"

#define pi             3.1415926535897931
#define SperA          31556926.0   /* seconds per year; 365.2422 days */

#define c_p_ICE        2009.0       /* J/(kg K)  specific heat capacity of ice */
#define rho_ICE        910.0        /* kg/(m^3)  density of ice */
#define k_ICE          2.10         /* J/(m K s) = W/(m K)  thermal conductivity of ice */
#define c_p_BRdefault  1000.0       /* J/(kg K)  specific heat capacity of bedrock */
#define rho_BRdefault  3300.0       /* kg/(m^3)  density of bedrock */
#define k_BRdefault    3.0          /* J/(m K s) = W/(m K)  thermal conductivity of bedrock */

#define H0             3000.0       /* m */
#define B0             1000.0       /* m */
#define Ts             223.15       /* m */
#define G              0.042        /* W/(m^2) */
#define phi            0.0125       /* K/m */

#define Nsum           30           /* number of terms in eigenfunction expansion; the exact
                                       solution is deliberately chosen to have finite expansion */


int exactK(const double t, const double z, double *TT, bool bedrockIsIce) {
  int k;
  bool belowB0;
  double ZZ, P, alpha, lambda, beta, gamma, XkSQR, Xk, theta, Ck, I1, I2, aH, bB, mI, mR;
  double c_p_BR, rho_BR, k_BR;
  /* following constants were produced by calling print_alpha_k(30) (below) */
  double alf[Nsum] = {3.350087528822397e-04, 1.114576827617396e-03, 1.953590840303518e-03,
                      2.684088585781064e-03, 3.371114869333445e-03, 4.189442265117592e-03,
                      5.008367405382524e-03, 5.696044031764593e-03, 6.425563506942886e-03,
                      7.264372872913219e-03, 8.044853066396166e-03, 8.714877612414516e-03,
                      9.493529164160654e-03, 1.033273985210279e-02, 1.106421822502108e-02,
                      1.175060460132703e-02, 1.256832682090360e-02, 1.338784224692084e-02,
                      1.407617951778051e-02, 1.480472324161026e-02, 1.564331999062109e-02,
                      1.642470780103220e-02, 1.709475346624607e-02, 1.787248418996684e-02,
                      1.871188358061674e-02, 1.944434477688470e-02, 2.013010181370026e-02,
                      2.094721145334310e-02, 2.176730968036079e-02, 2.245631776169424e-02};

  if (bedrockIsIce) {
    c_p_BR = c_p_ICE;
    rho_BR = rho_ICE;
    k_BR = k_ICE;
    for (k = 0; k < Nsum; k++) { /* overwrite alpha_k with ice-meets-ice values; see preprint */
      alf[k] = (2.0 * k + 1.0) * pi / (2.0 * (H0 + B0));
    }
  } else {
    c_p_BR = c_p_BRdefault;
    rho_BR = rho_BRdefault;
    k_BR = k_BRdefault;
  }
  if (z > H0) {
    *TT = Ts;
    return 0;
  }
  belowB0 = (z < -B0);

  ZZ = sqrt((rho_BR * c_p_BR * k_ICE) / (rho_ICE * c_p_ICE * k_BR));
  mI = (G / k_ICE) - phi;     mR = (G / k_BR) - phi;
  /* DEBUG: printf("ZZ = %10e, mI = %10e, mR = %10e\n", ZZ,mI,mR); */
  *TT = 0.0;
  for (k = Nsum-1; k >= 0; k--) {
    /* constants only having to do with eigenfunctions; theta = theta_k(z) is the
       normalized eigenfunction */ 
    alpha = alf[k];
    beta = ZZ * alpha;
    gamma = sin(alpha * H0) / cos(beta * B0);
    XkSQR = (rho_BR * c_p_BR * gamma * gamma * B0 + rho_ICE * c_p_ICE * H0) / 2.0;
    Xk = sqrt(XkSQR);
    theta = ( (z >= 0) ? sin(alpha * (H0 - z)) : gamma * cos(beta * (B0 + z)) ) / Xk;
    lambda = (k_ICE * alpha * alpha) / (rho_ICE * c_p_ICE);
    /* DEBUG: printf("k = %3d:  alpha = %10e, Xk = %10e, theta = %10e, lambda = %10e,\n",
           k,alpha,Xk,theta,lambda); */
    /* constants involved in computing the expansion coefficients */
    aH = alpha * H0;            bB = beta * B0;
    I1 = - mI * (sin(aH) - aH * cos(aH)) / (alpha * alpha);
    I2 = mR * (cos(bB) - 1.0 + bB * sin(bB)) / (beta * beta)
         - (B0 * mR + H0 * mI) * sin(bB) / beta;
    Ck = (rho_ICE * c_p_ICE * I1 + rho_BR * c_p_BR * gamma * I2) / Xk;
    /* add the term to the expansion */
    *TT += Ck * exp(- lambda * t) * theta;
    /* DEBUG: printf("          I1 = %10e, I2 = %10e, Ck = %10e, term = %10f\n",
           I1,I2,Ck, Ck * exp(- lambda * t) * theta ); */
  }
  P = (z >= 0) ? (z / k_ICE) - (H0 / k_ICE) : (z / k_BR) - (H0 / k_ICE);
  *TT += Ts - G * P;

  return ((belowB0) ? 1 : 0);

}


#define COMPUTE_ALPHA 0
#if COMPUTE_ALPHA

#define ALPHA_RELTOL   1.0e-14
#define ITER_MAXED_OUT 999

/* parameters needed for root problem: */
struct coscross_params {
  double Afrac, HZBsum, HZBdiff;
};

/* the root problem is to make this function zero: */
double coscross(double alpha, void *params) {
  struct coscross_params *p = (struct coscross_params *) params;
  return cos(p->HZBsum * alpha) - p->Afrac * cos(p->HZBdiff * alpha);
}

/* compute the first N roots alpha_k of the equation
     ((A-1)/(A+1)) cos((H - Z B) alpha) = cos((H + Z B) alpha)
where H and B are heights and A, Z are defined in terms of material 
constants */
int print_alpha_k(const int N) {
  int status, iter, k, max_iter = 200;
  double Z, A;
  double alpha, alpha_lo, alpha_hi, temp_lo;
  const gsl_root_fsolver_type *solvT;
  gsl_root_fsolver *solv;
  gsl_function F;
  struct coscross_params params;
  
  Z = sqrt((rho_BR * c_p_BR * k_ICE) / (rho_ICE * c_p_ICE * k_BR));
  A = (k_BR / k_ICE) * Z;
  params.Afrac   = (A - 1.0) / (A + 1.0);     
  params.HZBsum  = H0 + Z * B0;
  params.HZBdiff = H0 - Z * B0;
     
  F.function = &coscross;
  F.params = &params;
  solvT = gsl_root_fsolver_brent;  // faster than bisection but still bracketing
  solv = gsl_root_fsolver_alloc(solvT);

  for (k = 0; k < N; k++) {
    // these numbers bracket exactly one solution
    alpha_lo = (double(k) * pi) / params.HZBsum;
    alpha_hi = (double(k + 1) * pi) / params.HZBsum;
    gsl_root_fsolver_set(solv, &F, alpha_lo, alpha_hi);
     
    iter = 0;
    do {
      iter++;
      status = gsl_root_fsolver_iterate(solv);
      alpha = gsl_root_fsolver_root(solv);
      alpha_lo = gsl_root_fsolver_x_lower(solv);
      alpha_hi = gsl_root_fsolver_x_upper(solv);
      temp_lo = (alpha_lo > 0) ? alpha_lo : (alpha_hi/2.0);
      status = gsl_root_test_interval(temp_lo, alpha_hi, 0, ALPHA_RELTOL);
    } while ((status == GSL_CONTINUE) && (iter < max_iter));
    if (iter >= max_iter) {
      printf("!!!ERROR: root finding iteration reached maximum iterations; QUITING!\n");
      return ITER_MAXED_OUT;
    }
    printf("%19.15e,\n",alpha);
    /* DEBUG: printf("%19.15e  (in orig bracket [%19.15e,%19.15e])\n",alpha,
              (double(k) * pi) / params.HZBsum, (double(k+1) * pi) / params.HZBsum); */
  }
  
  gsl_root_fsolver_free(solv);
  return status;
}
#endif /* COMPUTE_ALPHA */

\end{verbatim}
\centerline{\rule{4.6in}{0.2mm}}
\bigskip

\normalsize

\end{document}